\documentclass[ aps, showpacs, showkeys, nofootinbib, floatfix, superscriptaddress]{revtex4}

\usepackage{amsfonts}

\usepackage{amssymb}
\usepackage{amsmath}
\usepackage{graphicx}

\begin{document}

\title{Does accelerating universe indicates  Brans-Dicke theory?}

 \author{Jianbo Lu}
 \email{lvjianbo819@163.com}
 \affiliation{Department of Physics, Liaoning Normal University, Dalian 116029, P. R. China}
 \author{Weiping Wang}
 \affiliation{Department of Physics, Liaoning Normal University, Dalian 116029, P. R. China}
 \author{Lixin Xu}
 \affiliation{Korea Astronomy and Space Science Institute, Daejon  305-348,
  Korea}
 \affiliation{School of Physics and Optoelectronic Technology, Dalian University of Technology,
 Dalian, 116024, P. R. China}
 \author{Yabo Wu}
 \affiliation{Department of Physics, Liaoning Normal University, Dalian 116029, P. R. China}

\begin{abstract}
 The evolution of universe in Brans-Dicke (BD) theory is discussed in this paper.
  Considering a parameterized scenario for BD scalar field
$\phi=\phi_{0}a^{\alpha}$ which plays the role of gravitational
¡±constant¡± $G$, we apply  the Markov Chain Monte Carlo method to
investigate a global constraints on BD theory with a
self-interacting potential according to the current observational
data:  the Union2 dataset of type supernovae Ia (SNIa),
 the high-redshift Gamma-Ray Bursts (GRBs) data,  the observational Hubble
data (OHD), the cluster X-ray gas mass fraction, the baryon acoustic
oscillation (BAO), and the cosmic microwave background (CMB) data.
It is shown that an expanded universe from deceleration to
acceleration is
 given in this theory, and the constraint results of dimensionless
 matter density  $\Omega_{0m}$ and parameter $\alpha$ are,
 $\Omega_{0m}=0.286^{+0.037+0.050}_{-0.039-0.047}$ and $\alpha=0.0046^{+0.0149+0.0171}_{-0.0171-0.0206}$
 which is consistent with the result of current
experiment exploration, $\mid\alpha\mid \leq 0.132124$. In addition,
we use the  geometrical diagnostic method, jerk parameter $j$, to
distinguish the BD theory and  the cosmological constant model in
Einstein's theory of general relativity.
\end{abstract}

\pacs{98.80.-k}

\keywords{Accelerating universe; Brans-Dicke theory; observational
constraint.}

\maketitle

\section{$\text{Introduction}$}

{The observation of the supernovae of type Ia \cite{SNIa,SNIa1}
provides the evidence that the universe is undergoing accelerated
expansion. Combining the observations from Cosmic Background
Radiation \cite{CMB} and lager scale structure \cite{LSS}, one
concludes that the universe at present is dominated by 70\% exotic
component.  An interpretation to the accelerating universe can just
be obtained by introducing an exotic fluid in the theory of general
relativity. Considering the standard cosmological model this unknown
component is called as dark energy with owning a character of
negative pressure to push the universe to an accelerated expansion.
And the popular dark energy model is the positive tiny cosmological
constant (CC), though it suffers the so-called fine tuning and
cosmic coincidence problems. However, in 2$\sigma$ confidence level,
it fits the observations very well \cite{lcdm2sigma}. If the
cosmological constant is not a real constant but is time variable,
the fine tuning and cosmic coincidence problems can be removed. In
fact, this possibility was considered in the past years
\cite{DEmodels,DEmodels1,DEmodels2,DEmodels3,DEmodels4,DEmodels5,DEmodels6,DEmodels7,DEmodels7a,DEmodels8,DEmodels9,DEmodels10,DEmodels11},
such as quintessence,
 Chaplygin gas, holographic, agegraphic dark energy, etc.
Also, the accelerating universe is related to the modification of
the gravity theory  on large distances, such as $f(R)$ modified
gravity theory \cite{fR} and higher dimensional theory
\cite{higherD,higherD1}, etc. Scalar-tensor modified gravity
theories have recently attracted much attention, in part because
they emerge naturally as the low-energy limit of  string theory as a
result of the dilaton coupling with gravitons
\cite{1003.1725,1003.1725-25-BDstring1,1003.1725-27-BDstring2,0205109,1005.0868-2,1005.0868-2-1}.
In addition, it has been argued that in the early universe, gravity
might obey a scalar tensor type theories rather than general
relativity (GR) \cite{0205109}, and they are also important for
cosmological inflation models
\cite{1003.1725-25-BDstring1,1003.1725-28-BDinflation,1003.1725-28-BDinflation1,1003.1725-28-BDinflation2},
since they have effectively solved the problems of inflation
\cite{0205109}, where the end of inflation can be brought by
nucleation without considering the fine-tuning cosmological
parameters.

\section{$\text{Brans-Dicke theory}$}

 Brans-Dicke (BD) theory \cite{BDtheory,BDtheory1,BDtheory2} of gravity is a simple but very important one among the
scalar tensor theories, which is apparently compatible with Mach's
principle \cite{0411066}. The generalized BD theory is an extension
of the original BD theory by considering  coupling parameter
$\omega$ as a function of the scalar field
\cite{1006.2609-8,1006.2609-8-1,1006.2609-8-2,1006.2609-8-3}, and an
accelerating universe can be obtained when coupling parameter
$\omega$ varies with time \cite{GBD-acceleration,GBD-acceleration1}.
In the framework of generalized BD theory, the action is described
as (choosing the speed of light $c=1$)
\begin{equation}
S=d^{4}x\sqrt{-g}[\phi R
-\frac{\omega(\phi)}{\phi}\phi^{,\alpha}\phi_{,\alpha}-V(\phi)+L_{m}].
\end{equation}
where $L_{m}$ is the matter Lagrangian, $\phi$ is the BD scalar
field which is non-minimally coupled to gravity, $V (\phi)$ is the
self-interacting potential for the BD scalar field, $\omega(\phi)$
being a function of $\phi$ is generalized version of the
dimensionless BD coupling parameter $\omega$, and $\phi$ plays the
role of the gravitational constant $G$ which is related to the
inverse of  $\phi$.

 The gravitational field equation
derived from above action  by varying the action with respect to the
metric is
\begin{equation}
 G_{\mu\nu}=\frac{8\pi}{\phi}T^{m}_{\mu\nu}+\frac{\omega(\phi)}{\phi^{2}}
 [\phi_{,\mu}\phi_{,\nu}-\frac{1}{2}g_{\mu\nu}\phi_{,\alpha}\phi^{,\alpha}]
 +\frac{1}{\phi}[\phi_{,\mu;\nu}-g_{\mu\nu}\Box
 \phi]-\frac{V(\phi)}{2\phi}g_{\mu\nu}
 \end{equation}
with
 \begin{equation}
 \Box \phi=\frac{8 \pi T}{3+2\omega(\phi)}-\frac{1}{3+2\omega(\phi)}[2V(\phi)-\phi\frac{d
 V(\phi)}{d \phi}]-\frac{\frac{d \omega(\phi)}{d \phi}}{3+2\omega(\phi)}\phi_{,\mu}\phi^{,\mu},
\end{equation}
where $G_{\mu \nu}$ is the Einstein tensor,  $T_{\mu \nu}^{m}$
denotes the energy-momentum tensor of matter, and
$T=T_{\mu\nu}^{m}g^{\mu\nu}$. Considering Robertson-Walker universe
 \begin{equation}
ds^{2}=-dt^{2}+a^{2}(t)[\frac{dr^{2}}{1-kr^{2}}+r^{2}(d\theta^{2}+sin^{2}\theta,
d\phi^{2})],
\end{equation}
where $k$ denotes the spacial curvature with $k=-1, 0, +1$
corresponding to open, flat and closed universe, respectively. The
modified Friedmann equation and wave equation for the generalized BD
scalar field $\phi$ are given as \cite{Hvbd,Hvbd1,Hvbd2},
\begin{equation}
H^{2}+\frac{k}{a^{2}}=\frac{8 \pi
\rho_{m}}{3\phi}-H\frac{\dot{\phi}}{\phi}+\frac{\omega(\phi)}{6}(\frac{\dot{\phi}}{\phi})^{2}+\frac{V(\phi)}{6\phi}
\equiv \frac{8 \pi}{3} \rho_{eff},\label{H1-GBD}
\end{equation}
\begin{equation}
2\frac{\ddot{a}}{a}+H^{2}+\frac{k}{a^{2}}=-\frac{8 \pi
p}{\phi}-\frac{\omega(\phi)}{2}(\frac{\dot{\phi}}{\phi})^{2}-2H\frac{\dot{\phi}}{\phi}
-\frac{\ddot{\phi}}{\phi}+\frac{V(\phi)}{2\phi} \equiv-8\pi
p_{eff},\label{H2-GBD}
\end{equation}
and
\begin{equation}
\ddot{\phi}+3H\dot{\phi}=\frac{8 \pi
(\rho_{m}-3p)}{3+2\omega(\phi)}+\frac{1}{3+2\omega(\phi)}[2V(\phi)-\phi\frac{d
V(\phi)}{d
\phi}]-\frac{\frac{d\omega(\phi)}{d\phi}}{3+2\omega(\phi)},\label{phi-GBD}
\end{equation}
where $\rho_{m}$ and $p$ denote the energy density and pressure for
the matter, $\rho_{eff}$ and  $p_{eff}$ are effective energy density
and pressure for the combination of matter and BD scalar field.
Obviously, the conservation equation of effective fluid is
satisfied, $\dot{\rho}_{eff}+3H(\rho_{eff}+3p_{eff})=0$. BD theory
as a leading alternative to Einstein's theory of GR, it can be
obtained by considering  $\omega(\phi)=\omega=$constant in above
generalized form.   Also, GR as  a special case in BD theory, the
theory can be  approximately reduced to GR when take constant
coupling parameter $\omega\rightarrow \infty$ and scalar function
$\phi=$constant.

\section{$\text{Cosmic constraints on Brans-Dicke theory}$}

 It has been shown that the BD theory is effective  for interpreting
the cosmic acceleration by considering a non-minimal coupling
between the gravitational term and the BD scalar field
\cite{Hvbd,Hvbd1,Hvbd2}. In this work, we investigate the global
cosmological constraints on BD theory. In order to solve above
 equations of motion the scale factor
$a=a_{0}t^{\beta}$ is considered in keeping with the recent
observations \cite{BDa-power}. Then the BD scalar field $\phi$ is
solved as $\phi=\phi_{0}t^{\gamma}$ as a function of time
\cite{BDa-power}. Here $\beta$ and $\gamma$ are two constant
parameters. Also, these two power-law solutions of the scale factor
$a$ and the BD scalar field $\phi$ can be found in other references
\cite{BD-phi-power,BD-phi-power1,BD-phi-power2,BD-phi-power3,BD-phi-power4,BD-phi-power5,BD-phi-power6},
which are consistent with the evolution of expanded
  universe. Inspired by these papers,
for calculation we consider a parameterized power-law form of
Brans-Dicke scalar field, $\phi=\phi_{0}a^{\alpha}$  which can be
easily  given by above two solutions of the scale factor $a$ and the
scalar field $\phi$. Then the Friedmann equation in BD theory with a
flat geometry $k=0$  is derived as,
\begin{eqnarray}
&H^{2}&=\frac{\rho_{m}}{3\phi}-\alpha
H^{2}+\frac{\omega}{6}\alpha^{2}H^{2}-\frac{V(\phi)}{6\phi}\nonumber\\
&&=\frac{2}{(6+6\alpha-\omega\alpha^{2})\phi_{0}}[\rho_{0m}a^{-(3+\alpha)}-\frac{\phi_{0}V(\phi)}{2\phi}].
\end{eqnarray}
 Defining
$\frac{2}{6+6\alpha-\omega\alpha^{2}}\frac{1}{\phi_{0}}=\frac{8 \pi
G_{0}}{3}$, then one has
\begin{eqnarray}
&H^{2}&=H^{2}_{0}\Omega_{0m}a^{-(3+\alpha)}-\frac{8 \pi
G_{0}}{3}\frac{\phi_{0}V(\phi)}{2\phi}\nonumber\\
&&=H^{2}_{0}\Omega_{0m}a^{-(3+\alpha)}-\frac{4 \pi
G_{0}}{3}a^{-\alpha}V(\phi), \label{H-parameterizedBD}
\end{eqnarray}
with $\Omega_{0m}\equiv \frac{8 \pi G_{0}\rho_{0m}}{3H^{2}_{0}}$.
 In addition,  for a flat
universe  according to  Eqs. (\ref{H1-GBD}), (\ref{H2-GBD}),
(\ref{phi-GBD}), and using the equation of state (EOS) for matter
$w_{m}=p/\rho_{m}=0$ and the conservation equation
$\rho_{m}=\rho_{0m}a^{-3(1+w_{m})}=\rho_{0m}a^{-3}$, the Friedmann
equation in BD theory can also be  expressed as \cite{BD-potential}
\begin{eqnarray}
&H^{2}&=Aa^{-3-\alpha}+Ba^{-\alpha\frac{2[(1+\omega)\alpha-1]}{2+\alpha}}\nonumber\\
&&=Aa^{-3-\alpha}+B a^{n},\label{H-BDpotential}
\end{eqnarray}
with  $n= \frac{-2\alpha[(1+\omega)\alpha-1]}{2+\alpha}$, here $A$,
$B$ are constant parameters. Comparing Eqs.
(\ref{H-parameterizedBD}) and (\ref{H-BDpotential}), one has
$A=H_{0}^{2}\Omega_{0m}$, and  the second term plays the role of
potential. Thus Friedmann equation is written as,
\begin{equation}
H^{2}=H^{2}_{0}\Omega_{0m}a^{-(3+\alpha)}+B a^{n}.
\end{equation}
If interpret the constant parameter $B$ in the second term as the
current value of energy density for an "new" component including the
BD potential,  the above equation becomes
\begin{equation}
H^{2}=H^{2}_{0}\Omega_{0m}a^{-(3+\alpha)}+H^{2}_{0}\Omega_{0x}a^{n}
\label{H-BDV-phin}
\end{equation}
 with writing  $B=H^{2}_{0}\Omega_{0x}$.
According to  Eq. (\ref{H-BDV-phin}) and considering the index
$n=-3(1+w_{x})$, it is obtained that the equation of state is
expressed as, $w_{x}=-1-\frac{n}{3}$ for this component.

\begin{figure}[ht]
  \includegraphics[width=11.5cm]{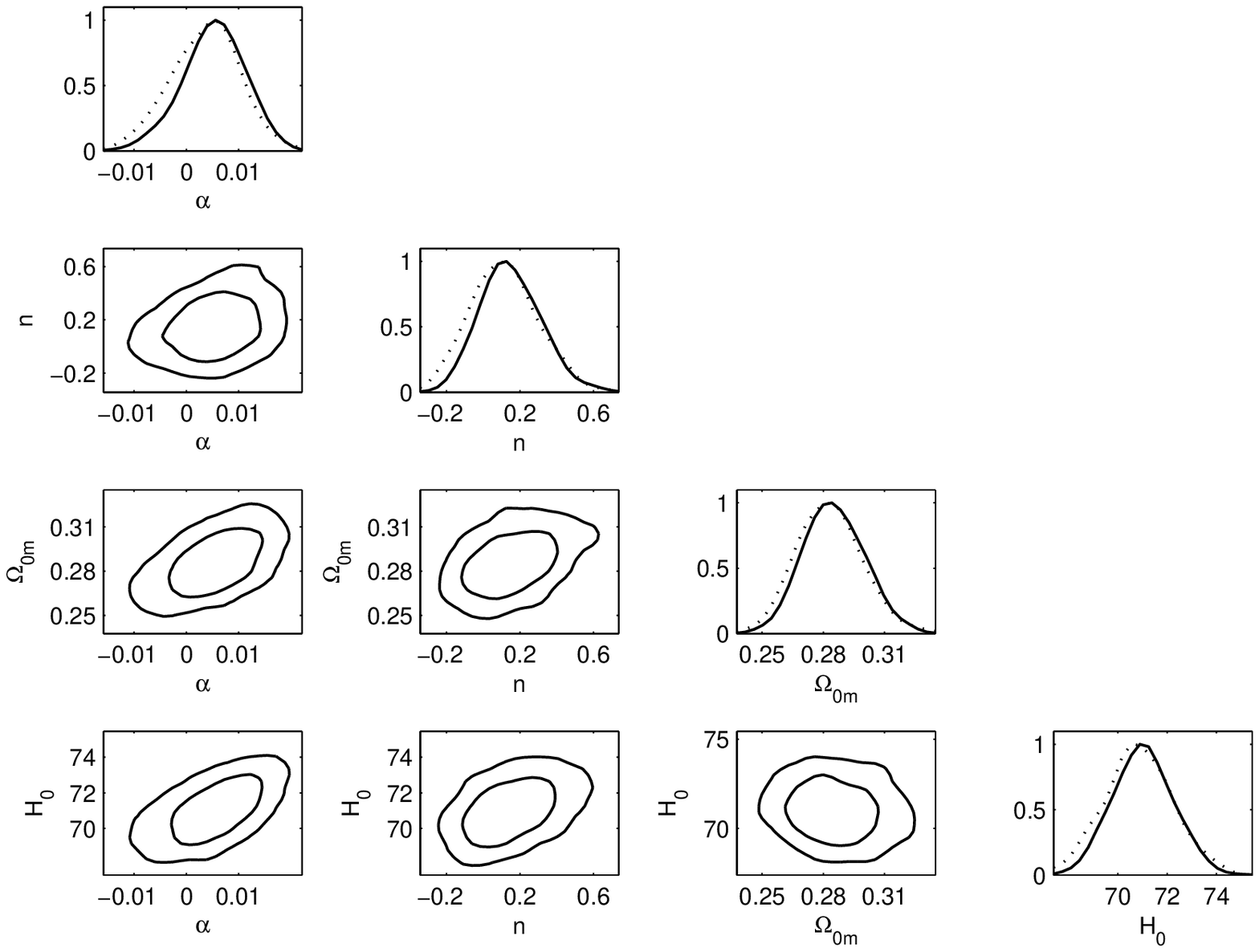}~
  \includegraphics[width=7cm]{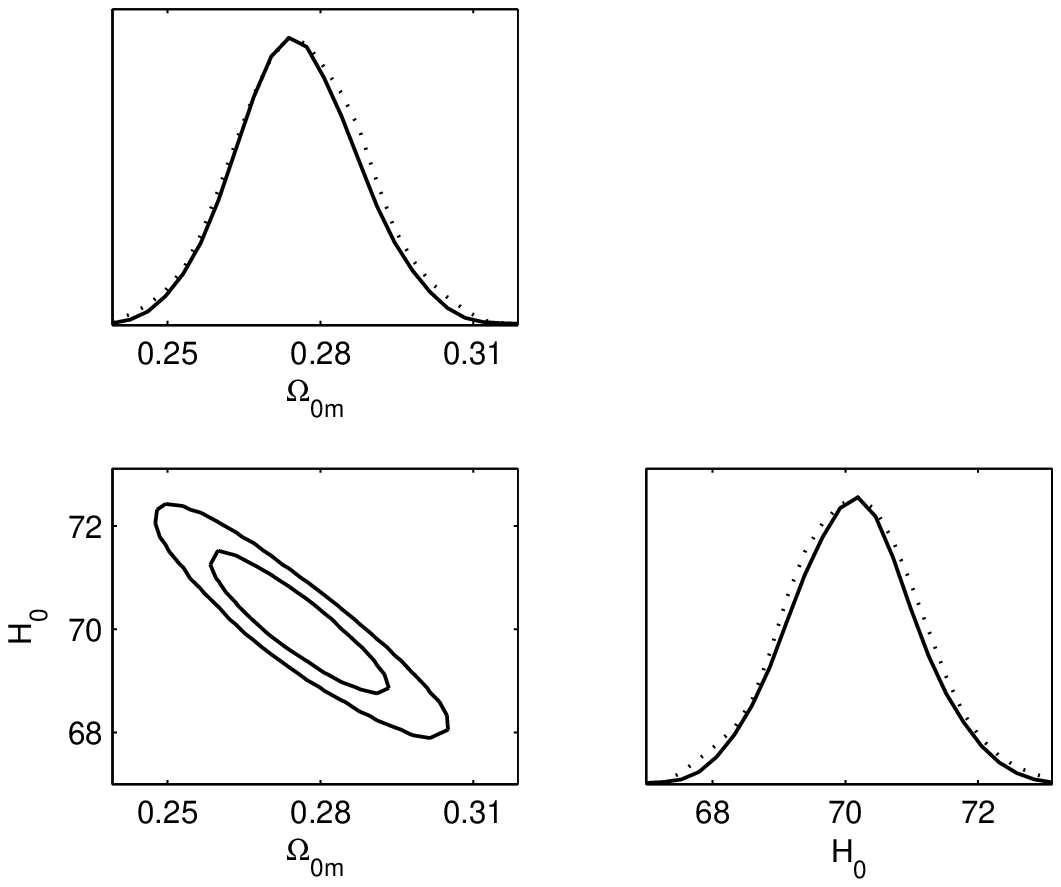}\\
  \caption{The 2-D contours with  $1\sigma, 2\sigma$ confidence levels and 1-D
  distribution of model parameters  in the flat BD (left) and CC (right) model.
  Solid lines are mean likelihoods
  of samples, and dotted  lines are marginalized probabilities for 1D
  distribution.}\label{figure-ab-616+4-VG-v-BD-phin}
\end{figure}

In the following we apply the Markov Chain Monte Carlo (MCMC) method
to investigate the global constraints on above BD scenario. For the
used observational data, we consider 557 Union2 dataset of type
supernovae Ia (SNIa) \cite{557Union2},   59 high-redshift Gamma-Ray
Bursts (GRBs) data \cite{GRBs30},  observational Hubble data (OHD)
\cite{OHD}, X-ray gas mass fraction in  cluster \cite{ref:07060033},
baryon acoustic oscillation (BAO) \cite{ref:Percival2}, and cosmic
microwave background (CMB) data \cite{7ywmap}, and for the analysis
method please see appendix. In our joint analysis, the MCMC code is
based on the publicly available CosmoMC package \cite{ref:MCMC} and
the modified CosmoMC package
\cite{ref:0409574,ref:07060033,ref:modifiedMCMC}. The latter package
is about the constraint code  of X-ray cluster gas mass fraction. In
the calculation the baryon matter density is taken to be varied with
a tophat prior: $\Omega_{b}h^{2}\in [0.005,0.1]$. In addition, we
run 8 independent chains in the MCMC calculation, and to get the
converged results we test the convergence of the chains by typically
getting $R - 1$ to be less than 0.03.
 The total $\chi^{2}$ is expressed as,
 \begin{equation}
 \chi^{2}_{total}(p_{s})=\chi^{2}_{SNIa}+\chi^{2}_{GRBs}+\chi^{2}_{OHD}+\chi^{2}_{CBF}+\chi^{2}_{BAO}+\chi^{2}_{CMB},\label{chi2total-4dynamical}
 \end{equation}
 and the parameter vector reads
\begin{equation}
 p_s=\{\Omega_{b}h^2, \Omega_{c}h^2, \alpha,n\}.
 \end{equation}
Here the expression of  $\chi^{2}$ for each observation corresponds
to Eqs.(\ref{eq:chi2SN}), (\ref{chi2GRBs}),
 (\ref{chi2OHD}), (\ref{eq:chi2fgas}), (\ref{chi2-BAO}) and (\ref{chi2-CMB}),
 $\Omega_{b}$ and
$\Omega_{c}$ denote dimensionless energy density of baryon matter
and dark matter, respectively.
 Based on the  basic cosmological parameters $p_{s}$ we can  also obtain the
derived parameters $\Omega_{0m}=\Omega_{b}+\Omega_{c}$,
$\Omega_{0x}=1-\Omega_{0m}$, and the Hubble constant $H_0=100h$
km$\cdot$s $^{-1}\cdot$Mpc$^{-1}$.
 Using the currently observed data with the $\chi^{2}_{total }$ in Eq. (\ref{chi2total-4dynamical}),
  Fig. \ref{figure-ab-616+4-VG-v-BD-phin} (left) plots the 2-D contours with  $1\sigma, 2\sigma$
  confidence levels and 1-D
  distribution of model parameters  in the flat BD theory.
  Solid lines are mean likelihoods
 of samples, and dotted  lines are marginalized probabilities for 1D
 distribution  (for Gaussian distributions they should
 be the same).   And the
   calculation results  are listed
   in table \ref{tableVG-V-BD-mcmc-means} for the constraint on model parameters. In addition,
Table \ref{tableVG-V-BD-mcmc-bestfit} shows the  values for the
best-fit sample, and projections of the n-Dimensional $1\sigma$ and
$2\sigma$ confidence regions.  The n-D limits give  some idea of the
range of the posterior, and are much more conservative than the
marginalized limits \cite{ref:MCMC}. Often,  the best-fit results
are recommended.  According to the best fit values of model
parameter, it is shown
  the value of equation of state is,
  $w_{x}=-1-\frac{n}{3}\simeq-1.034$,
  which is in the phantom region with a small deviation. And according to the best fit values of  $\alpha=0.0046$ and $n=0.102$, the best fit value of
 coupling parameter,
 $\omega=\frac{-n(2+\alpha)+2\alpha(1-\alpha)}{2\alpha^{2}}=-4615.11$ is calculated,
 which is consistent with other results, such as $\omega<-40$
 constrained from the growth rate data \cite{arXiv:1012.5451}, and
 $\omega<-120$ or $\omega >97.8$ constrained from the WMAP CMB and the SDSS BAO
 data \cite{values-omega}.
 In addition, comparing with the literature \cite{Xu-BDH},
 where the Brans-Dicke  theory  is constrained from the recent
 observational data with introducing the holographic dark energy in
 universe in order to obtain an accelerating universe, it can be
 found that our results in this paper have a  larger value of parameter
 $\alpha$.
   And for the constraint on the dimensionless
 matter density $\Omega_{0m}$ in the BD theory, from Table \ref{tableVG-V-BD-mcmc-bestfit}
   one can see that it is consistent with
results from other analyses, such as the constraints on several
dynamical dark energy models and model independent scenarios   by
the recently observed data \cite{C-Omegam,C-Omegam1,C-Omegam2},
where the best fit value of $\Omega_{0m}$ is about 0.27. At last, it
may be said that the BD theory with a self-potential  is slightly
preferred by the current observational data because the quantity
$\chi^{2}_{min}$ measures the goodness of model fit, i.e. the less
value of $\chi^{2}_{min}$, the better model of agreeing with
observations. Therefore, it seems that the BD theory with a
self-potential  is better than the cosmological constant to
interpret an accelerating universe, and it is also significative to
discuss the evolution of cosmological quantities in BD theory.

\begin{table}[ht]
 \vspace*{-12pt}
 \begin{center}
 \begin{tabular}{c |  c|  c|  c|  c } \hline\hline
      & $\alpha$ &  n &  $\Omega_{0m}$ &   $H_{0}$   \\\hline
   BD & $0.0052(\pm0.0059)^{+0.0060+0.0111}_{-0.0112-0.0127}$
      & $0.149(\pm0.164)^{+0.165+0.347}_{-0.271-0.298}$
      & $0.285(\pm0.015)^{+0.015+0.030}_{-0.014-0.027}$
      & $70.028(\pm1.201)^{+1.211+2.363}_{-1.216-2.271}$\\\hline
   CC & ---&---& $0.272(\pm0.012)^{+0.013+0.026}_{-0.013-0.025}$ &
      $70.118(\pm0.913)^{+0.899+1.830}_{-0.903-1.753}$  \\\hline\hline
 \end{tabular}
 \end{center}
 \caption{  The means, standard deviations (the numerical results in brackets)
   and the marginalized limits for the model parameters from MCMC calculation, obtained by using
  SNIa Union2, GRBs, OHD, CBF, BAO, and CMB  data. CC denotes the cosmological
 constant (CC) model in the Einstein's general relativity.}\label{tableVG-V-BD-mcmc-means}
 \end{table}

\begin{table}[!htbp]
 \vspace*{-12pt}
 \begin{center}
 \begin{tabular}{c | c| c | c|c|c|c  } \hline\hline
  & $\chi_{min}^{2}$     &$\frac{\chi_{min}^{2}}{dof}$   &$\alpha$ &$n$ &$\Omega_{0m}$ & $H_{0}$     \\\hline
  BD   & 619.466  &0.918
  &$0.0046^{+0.0149+0.0171}_{-0.0171-0.0206}$
  &$0.102^{+0.496+0.606}_{-0.382-0.479}$
  & $0.286^{+0.037+0.050}_{-0.039-0.047}$
  &$70.537^{+3.766+4.293}_{-2.703-3.326}$\\\hline
   CC     & 619.840  & 0.917 &---&---& $0.274^{+0.030+0.043}_{-0.026-0.032}$
   &$70.255^{+2.127+2.790}_{-2.293-2.915}$\\\hline\hline
 \end{tabular}
 \end{center}
 \caption{The maximum likelihood values $\chi_{min}^{2}$, $\chi_{min}^{2}/dof$,
   the best fit model parameters,
 and the limits from the extremal values of  N-dimensional distribution from the MCMC calculation by using
 SNIa Union2,  GRBs, CBF, OHD, BAO, and CMB data,
 where $dof$ is degree of freedom of model, and its value
equals the number of observational data points minus the number of
parameters. }\label{tableVG-V-BD-mcmc-bestfit}
 \end{table}

Considering the value of parameter $\alpha$ is not arbitrary,  we
calculate its  values by using  other experiment exploration on
Newton$^{'}$s gravitational "constant" $G$. According to the
independent
 observations of Hulse-Taylor binary pulsar B1913 + 16 \cite{VG-constraints1,VG-constraints1a},
 and astereoseismological data from the pulsating white dwarf star G117-B15A \cite{VG-constraints2,VG-constraints2a},
  it is indicated by the current constraints  on the variation of $G$,
\begin{equation}
\mid\frac{\dot{G}}{G}\mid \leq 10^{-11}yr^{-1}.\label{VG}
\end{equation}
For BD theory and the parameterized scenario
$\phi=\phi_{0}a^{\alpha}$, it corresponds to
\begin{equation}
\mid\frac{\dot{\phi}}{\phi}\mid=\mid\alpha H\mid\leq
10^{-11}yr^{-1},
\end{equation}
i.e.,
\begin{equation}
\mid\alpha\mid  \leq \frac{1}{H} \times 10^{-11}yr^{-1}.
\end{equation}
Considering the current value of Hubble constant
$h=0.742^{+0.036}_{-0.036}$ \cite{H0prior} and taking the best fit
value of $H_{0}=74.2$km$\cdot$s $^{-1}\cdot$Mpc$^{-1}=2.40\times
10^{-18} s^{-1}=7.57 \times 10^{-11}yr^{-1}$, one obtains the bounds
on $\alpha$,
\begin{equation}
\mid\alpha\mid \leq 0.132124.
\end{equation}
It is easy to see that the value of constraint on parameter $\alpha$
with $2\sigma$ confidence level for BD theory  is under this bound,
i.e., it is lied in a physical significative region.

\begin{figure}[ht]
  \includegraphics[width=5cm]{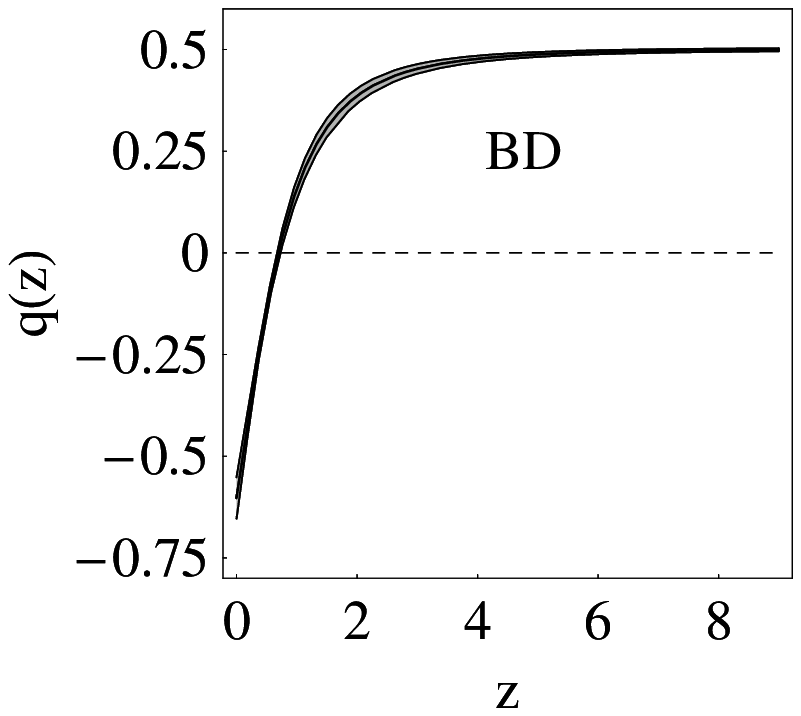}
  ~~ \includegraphics[width=5cm]{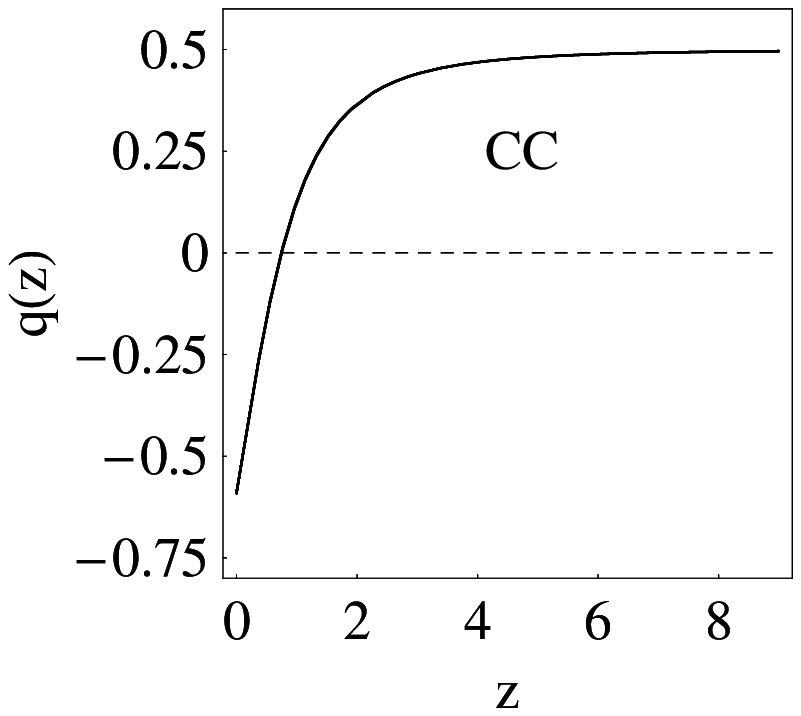}\\
  \caption{The evolution of $q(z)$  for BD and cosmological constant (CC) model.}\label{figureqz-VG-v-BD-phin}
\end{figure}

\begin{table}[ht]
 \vspace*{-12pt}
 \begin{center}
 \begin{tabular}{c | c | c } \hline\hline
   & $z_{T}$ $(1\sigma)$  & $q_{0}$ $(1\sigma)$  \\\hline
 BD    & $0.695^{+0.031}_{-0.028}$    & $-0.603^{+0.051}_{-0.050}$     \\\hline
 CC    &$0.744^{+0.007}_{-0.008}$   & $-0.589^{+0.003}_{-0.004}$   \\\hline\hline
 \end{tabular}
 \end{center}
 \caption{The  values of  transition redshift $z_{T}$,
 and  current deceleration parameter $q_{0}$ from MCMC calculation, obtained by using
 SNIa Union2,  GRBs, CBF, OHD, BAO, and CMB data. }\label{tableVG-V-BD-phin}
 \end{table}

Furthermore, we investigate the evolution of deceleration parameter
$q(z)=-\frac{\ddot{a}}{aH^{2}}=(1+z)\frac{1}{H}\frac{dH}{dz}-1$ in
this scenario. Considering the propagation of the errors for  $q(z)$
by the Fisher matrix analysis, the errors are evaluated by using the
covariance matrix $C_{ij}$ of the fitting parameters
\cite{cij45,cij46}, which is the inverse of the Fisher matrix and
given by
\begin{equation}
 (C_{ij}^{-1})=-\frac{\partial^{2}\ln L}{\partial p_{s_{i}}\partial p_{s_{j}}}
 =\frac{1}{2}\frac{\partial^{2}\chi^{2}(p_{s})}{\partial p_{s_{i}}\partial p_{s_{j}}},
\end{equation}
where $p_{s}$ is a set of parameters, and $\ln L$ is the logarithmic
likelihood function. The errors on a function $f = f(p_{s})$ in
terms of the variables $p_{s}$ are calculated by \cite{cij46,cij47}
\begin{equation}
 \sigma_{f}^{2}=\sum_{i}^{m}(\frac{\partial f}{\partial
  p_{s_{i}}})^{2}C_{ii}+2\sum_{i}^{m}\sum_{j=i+1}^{m}
  (\frac{\partial f}{\partial p_{s_{i}}})(\frac{\partial f}{\partial
  p_{s_{j}}})C_{ij},
\end{equation}
where $m$ is the number of parameters, and $f$ will be
 deceleration parameter $q(z; p_{s_{i}})$. The parameters $p_{s_{i}}$ respectively
represent $( \Omega_{b}h^{2},\Omega_{c}h^{2}, \alpha, n)$. In Fig.
\ref{figureqz-VG-v-BD-phin} (left) we plot the evolutions of $q(z)$
with errors by
\begin{equation}
q_{1\sigma}(z)=q(z)\mid_{p_{s}=\bar{p}_{s}}\pm \sigma _{q},
\end{equation}
here $\bar{p}_{s}$ are the best fit values of the constraint
parameters. From this figure it can be seen that a universe from
decelerated expansion to accelerated expansion is obtained. And the
values of transition redshift $z_{T}$ and current decelerating
parameter $q_{0}$ with $1\sigma$ confidence level are,
$z_{T}=0.692^{+0.028}_{-0.024}$ and
$q_{0}=-0.594^{+0.034}_{-0.035}$, as shown in table
\ref{tableVG-V-BD-phin}.

 For comparison  we also consider the observed constraints on
popular cosmological constant (CC), i.e. $\Lambda$CDM model in the
Einstein's theory of GR (or dubbed as standard cosmology). The
Friedmann equation for this case is expressed as
\begin{equation}
H^{2}=H^{2}_{0}\Omega_{0m}a^{-3}+H^{2}_{0}\Omega_{0\Lambda}.
\end{equation}
The calculation results on cosmological parameters are listed in
table \ref{tableVG-V-BD-mcmc-means} and
\ref{tableVG-V-BD-mcmc-bestfit}  for this model. In addition, for
the evolution of deceleration parameter $q(z)$ it is plotted in Fig.
\ref{figureqz-VG-v-BD-phin} (right), and the values of $z_{T}$ and
$q_{0}$ are listed in table \ref{tableVG-V-BD-phin}.

\section{$\text{jerk parameter geometrical  diagnostic for Brans-Dicke theory}$}

In the following we use a new geometrical diagnostic method, jerk
parameter $j$, to investigate discriminations between BD and CC
model. The jerk parameter is defined by scale factor $a$ and its
third derivative \cite{jerkvalue,jerkvalue1,jerkvalue2},
\begin{equation}
j \equiv
-\frac{1}{H^{3}}(\frac{\dot{\ddot{a}}}{a})=-[\frac{1}{2}(1+z)^{2}\frac{[H(z)^{2}]^{''}}{H(z)^{2}}
-(1+z)\frac{[H(z)^{2}]^{'}}{H(z)^{2}}+1].\label{jerk}
\end{equation}
 The use of the cosmic jerk parameter provides more
parameter space for geometrical  studies, and transitions between
phases of different cosmic acceleration are more naturally described
by models incorporating a cosmic jerk. In addition, for a dark
energy model  Eq. (\ref{jerk}) can be derived as
\begin{equation}
 j=-1-\frac{9}{2}w_{de}\Omega_{de}(1+w_{de})+\frac{3}{2}\Omega_{de}\frac{\dot{w}_{de}}{H},\label{jlcdm}
 \end{equation}
 where $\Omega_{de}$ denotes dimensionless energy density
 for dark energy. From Eq. (\ref{jlcdm}) it is easy to see that,
  for flat $\Lambda$CDM model ($w_{de}(z)=-1$), it has a constant jerk
 with $j(z) = -1$. Thus, it can provide
us with a simple, convenient approach  to distinguish and  search
departures for both  cosmological dynamic and kinematical models
from the cosmic concordance model, CC. Using Eq. (\ref{jerk}), we
plot the  evolution of jerk parameter $j(z)$ for BD model in Fig.
\ref{figurejz-VG-v-BD-phin} and compare it with cosmological
constant. According to  this figure we get the current values of
jerk parameter,
 $j_{0}=-1.002^{+0.003}_{-0.004}$.
 And it is shown that  at $1\sigma$ confidence level the BD theory
 can  not be distinguished  with cosmological constant model in standard cosmology
 according to the jerk parameter geometrical diagnostic.

\begin{figure}[ht]
   \includegraphics[width=6cm]{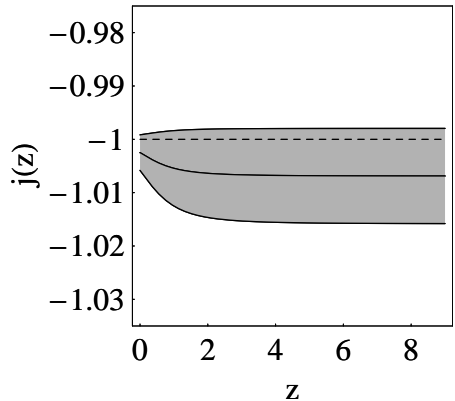}\\
  \caption{The evolution of jerk parameter $j(z)$  for BD model.}\label{figurejz-VG-v-BD-phin}
\end{figure}

\section{$\text{Conclusions}$}

Many observations and phenomena indicate that  the gravitational
theory of general relativity should be modified.  Scalar tensor (ST)
theory as a modified theory of gravity has been widely studied
\cite{BD-studies,BD-studies1,BD-studies2,BD-studies3,BD-studies4,BD-studies5,BD-studies6,BD-studies7,BD-studies8,BD-studies9}.
Considering the  ST theory can be well used to interpret the
inflation problem, in this paper Brans-Dicke gravitational theory as
an important one in ST theories is studied to interpret the late
accelerating universe. In BD theory gravitational "constant" $G$ is
described by a function of scalar field which couples to the Ricci
scalar $R$, so that the effective $G$ generally varies in time.
Concretely, with a parameterized scenario for BD scalar field $\phi$
which plays the role of the variable gravitational constant $G$, we
use the MCMC analysis method and the current observed data,
including 557 Union2 SNIa, 59 GRBs, OHD, cluster X-ray gas mass
fraction, BAO and CMB data, to constrain the BD theory with a
self-interacting potential. According to the constraint results on
deceleration parameter $q(z)$, the late accelerating universe can be
obtained in this theory. And for the parameter $\alpha$ that is
corresponded to the variable gravitational "constant" $G$ in
physics, its constraint value  is
 consistent with the result of  current experiment exploration,
$\mid\alpha\mid \leq 0.132124$. Considering  the application of BD
theory in cosmology, according to Eq. (\ref{H-BDV-phin}) since the
both energy densities  are dynamical, not a constant as in
cosmological constant model, it will not suffer the so-called cosmic
coincidence problem. In addition, relative to other geometrical
diagnostic methods, such as Om and statefinder parameters \{r,s\}
which have been widely  discussed  in many models
\cite{Om-rs,Om-rs1,Om-rs2,Om-rs3,Om-rs4,Om-rs5,Om-rs6}, in this
paper we use a new geometrical diagnostic method----jerk parameter
to Brans-Dicke theory, and it is shown that it can not be
distinguished with cosmological constant model. Obviously, the
perturbation theory could distinguish this modified gravity theory
of BD with the CC dark energy scenario  in Einstein's theory of
general relativity, which deserves to be studied in the future.

 \textbf{\ Acknowledgments }
 One of authors Jianbo Lu thanks Changjun Gao and Wei Fang for useful
 discussion in this work.
 The research work is supported by the National Natural Science Foundation
  (Grant No. 10875056) and NSF (10703001)  of P.R. China.

\appendix
\section{$\text{Observational data and cosmological constraint methods}$}\label{constraint-method}
 In this part we introduce the cosmological constraints methods and the current observed data used in this paper.
\subsection{Type Ia supernovae}

We use the 557 SNIa Union2 dataset, which includes $557$ SNIa
\cite{557Union2}. Following
\cite{ref:smallomega,ref:smallomega1,ref:smallomega2,ref:POLARSKI},
one can obtain the corresponding constraints by fitting the distance
modulus $\mu(z)$ as
\begin{equation}
\mu_{th}(z)=5\log_{10}[D_{L}(z)]+\frac{15}{4}\log_{10}\frac{G}{G_{0}}+\mu_{0},
\end{equation}
where  $G_{0}$ is the current value of  Newton's constant $G$. In
this expression $D_{L}(z)$ is the Hubble-free luminosity distance
$H_0 d_L(z)/c$, with $H_0$ the Hubble constant, defined through the
re-normalized quantity $h$ as $H_0 =100 h~{\rm km ~s}^{-1} {\rm
Mpc}^{-1}$,
 and
\begin{eqnarray}
d_L(z)&=&\frac{c(1+z)}{\sqrt{|\Omega_k|}}sinn[\sqrt{|\Omega_k|}\int_0^z\frac{dz'}{H(z')}], \nonumber\\
 &&\mu_{0}=5log_{10}(\frac{H_{0}^{-1}}{Mpc})+25=42.38-5log_{10}h.\nonumber
\end{eqnarray}
where $sinnn(\sqrt{|\Omega_k|}x)$ respectively denotes
$\sin(\sqrt{|\Omega_k|}x)$, $\sqrt{|\Omega_k|}x$,
$\sinh(\sqrt{|\Omega_k|}x)$ for $\Omega_k<0$, $\Omega_k=0$ and
$\Omega_k>0$.
 Additionally, the observed distance moduli $\mu_{obs}(z_i)$ of SNIa at $z_i$ is
\begin{equation}
\mu_{obs}(z_i) = m_{obs}(z_i)-M,
\end{equation}
where $M$ is their absolute magnitudes.

For using SNIa data, theoretical model parameters  $p_s$ can be
determined by a likelihood analysis, based on the calculation of
\begin{eqnarray}
\chi^2(p_s,M^{\prime})\equiv \sum_{SNIa}\frac{\left\{
\mu_{obs}(z_i)-\mu_{th}(p_s,z_i)\right\}^2} {\sigma_i^2}
=\sum_{SNIa}\frac{\left\{ 5 \log_{10}[D_L(p_s,z_i)] - m_{obs}(z_i) +
M^{\prime} \right\}^2} {\sigma_i^2}, \ \ \ \ \label{eq:chi2}
\end{eqnarray}
where $M^{\prime}\equiv\mu_0+M$ is a nuisance parameter which
includes the absolute magnitude and the parameter $h$. The nuisance
  parameter $M^{\prime}$ can be marginalized over
analytically
\cite{ref:SNchi2,ref:SNchi21,ref:SNchi22,ref:SNchi23,ref:SNchi24,ref:SNchi25,ref:SNchi26}
as
\begin{equation}
\bar{\chi}^2(p_s) = -2 \ln \int_{-\infty}^{+\infty}\exp \left[
-\frac{1}{2} \chi^2(p_s,M^{\prime}) \right] dM^{\prime},\nonumber
\label{eq:chi2marg}
\end{equation}
resulting to
\begin{equation}
\bar{\chi}^2 =  A - \frac{B^2}{C} + \ln \left( \frac{C}{2\pi}\right)
, \label{eq:chi2mar}
\end{equation}
with
\begin{eqnarray}
&&A=\sum_{SNIa} \frac {\left\{5\log_{10}
[D_L(p_s,z_i)]-m_{obs}(z_i)\right\}^2}{\sigma_i^2},\nonumber\\
&& B=\sum_{SNIa} \frac {5
\log_{10}[D_L(p_s,z_i)]-m_{obs}(z_i)}{\sigma_i^2},\nonumber
\\
&& C=\sum_{SNIa} \frac {1}{\sigma_i^2}\nonumber.
\end{eqnarray}
Relation (\ref{eq:chi2}) has a minimum at the nuisance parameter
value $M^{\prime}=B/C$, which contains information of the values of
$h$ and $M$. Therefore, one can extract the values of $h$ and $M$
provided the knowledge of one of them. Finally, note that the
expression
\begin{equation}
\chi^2_{SNIa}(p_s)=A-(B^2/C),\label{eq:chi2SN}\nonumber
\end{equation}
which coincides to (\ref{eq:chi2mar}) up to a constant, is often
used in the likelihood analysis
\cite{ref:smallomega,ref:smallomega1,ref:SNchi2,ref:SNchi21,ref:SNchi22,ref:SNchi23},
and thus in this case the results will not be affected by a flat
$M^{\prime}$ distribution. For minimizing $\chi^{2}_{SNIa}(p_{s})$
to perform a constraint on cosmological parameters, it is equivalent
to maximizing the likelihood
\begin{equation}
L(p_{s})\propto \exp[\frac{-\chi^{2}(p_{s})}{2}].\label{Lsn}
\end{equation}

\subsection{high-redshift Gamma-Ray Bursts data}

 The GRBs data can be observed at higher redshift than
SNIa. The currently observed reshift range of GRBs is at $0.1
\lesssim z \lesssim 9$. Therefore, the GRBs data can be viewed as an
excellent complement to SNIa data and would provide more information
at high redshift. When several empirical relations of the GRBs are
proposed, these indicators have motivated the authors make use of
the GRBs as cosmological standard candles at high redshift. However,
the fact that there are not sufficient low reshift GRBs leads that
the calibration of GRB relations is dependent on the cosmological
model, namely, the circularity problem. One of methods to solve the
circularity problem is the calibration of GRB relations are
performed by the use of a sample of SNIa at low redshift in the
cosmology-independent way \cite{GRBs48}. Here, the GRBs data we used
consists of 59 GRB samples with a redshift range of $1.4 \lesssim z
\lesssim 9$ obtained in \cite{GRBs30}. These 59 GRBs are calibrated
by utilizing the newly released 557 Uion2 SNIa  and the isotropic
energy-peak spectral energy ($E_{iso}$- $E_{p,i}$) relation (i.e.
Amati relation) \cite{GRBs49}.

The $\chi^{2}_{GRBs}$ takes the same form as $\chi^{2}_{SNIa}$
\begin{equation}
\chi^{2}_{GRBs}(p_{s},\mu_{0})=\sum_{i=1}^{59}\frac{[\mu_{obs}(z_{i}
-\mu_{th}(z_{i};p_{s},\mu_{0})]^{2}}{\sigma_{i}^{2}}.\label{chi2GRBs}
\end{equation}
The same method are used to deal with the nuisance parameter
$\mu_{0}$ as shown in the description of $\chi^{2}_{SNIa}$ above.

\subsection{Observational Hubble data}

The observational Hubble data \cite{ohdzhang} are based on
differential ages of the galaxies. In \cite{ref:JVS2003}, Jimenez
{\it et al.} obtained an independent estimate for the Hubble
parameter using the method developed in \cite{ref:JL2002}, and used
it to constrain the cosmological models. The Hubble parameter
depending on the differential ages as a function of redshift $z$ can
be written in the form of
\begin{equation}
H(z)=-\frac{1}{1+z}\frac{dz}{dt}.
\end{equation}
So, once $dz/dt$ is known, $H(z)$ is obtained directly. By using the
differential ages of passively-evolving galaxies from the Gemini
Deep Deep Survey (GDDS) \cite{ref:GDDS} and archival data
\cite{ref:archive1,ref:archive2,ref:archive3,ref:archive4,ref:archive5,ref:archive6},
Simon {\it et al.} obtained several values of $H(z)$  at  different
redshift \cite{OHD}. The twelve observational Hubble data  (redshift
interval  $0\lesssim z \lesssim 1.8$) from
\cite{12Hubbledata,12Hubbledata1,H0prior} are list in Table
\ref{table-12Hubbledata}.
\begin{table}[ht]
\begin{center}
\begin{tabular}{c|llllllllllll}
\hline\hline
 $z$ &\ 0 & 0.1 & 0.17 & 0.27 & 0.4 & 0.48 & 0.88 & 0.9 & 1.30 & 1.43 & 1.53 & 1.75  \\ \hline
 $H(z)\ ({\rm km~s^{-1}\,Mpc^{-1})}$ &\ 74.2 & 69 & 83 & 77 & 95 & 97 & 90 & 117 & 168 & 177 & 140 & 202  \\ \hline
 $1 \sigma$ uncertainty &\ $\pm 3.6$ & $\pm 12$ & $\pm 8$ & $\pm 14$ & $\pm 17$ & $\pm 60$ & $\pm 40$
 & $\pm 23$ & $\pm 17$ & $\pm 18$ & $\pm 14$ & $\pm 40$ \\
\hline\hline
\end{tabular}
\end{center}
\caption{\label{table-12Hubbledata} The observational $H(z)$
data~\cite{12Hubbledata,12Hubbledata1,H0prior}.}
\end{table}
In addition, in \cite{3Hubbledata} the authors take the BAO scale as
a standard ruler in the radial direction, and obtain three more
additional data: $H(z=0.24)=79.69\pm2.32, H(z=0.34)=83.8\pm2.96,$
and $H(z=0.43)=86.45\pm3.27$.

 The best fit values of the model parameters from observational Hubble data  are determined by minimizing \cite{chi2hub,chi2hub1,chi2hub2}
 \begin{equation}
 \chi_{OHD}^2(H_{0},p_{s})=\sum_{i=1}^{15} \frac{[H_{th}(H_{0},p_{s};z_i)-H_{obs}(z_i)]^2}{\sigma^2(z_i)},\label{chi2OHD}
 \end{equation}
 where  $H_{th}$ is the predicted value for the Hubble parameter, $H_{obs}$ is the observed value, $\sigma(z_i)$ is the standard
 deviation measurement uncertainty, and the summation is over the $15$ observational Hubble data points at redshifts $z_i$.

\subsection{The X-ray gas mass fraction}
The X-ray gas mass fraction, $f_{gas}$, is defined as the ratio of
the X-ray gas mass to the total mass of a cluster, which is
approximately independent on the redshift for the hot ($kT \gtrsim
5keV$), dynamically relaxed clusters at the radii larger than the
innermost core $r_{2500}$. As inspected in \cite{ref:07060033}, the
$\Lambda$CDM model is very favored and has been chosen as the
reference cosmology. The model fitted to the reference $\Lambda$CDM
data is presented as \cite{ref:07060033}
\begin{eqnarray}
&&f_{gas}^{\Lambda CDM}(z)=\frac{K A \gamma
b(z)}{1+s(z)}\left(\frac{\Omega_b}{\Omega_{0m}}\right)
\left[\frac{D_A^{\Lambda CDM}(z)}{D_A(z)}\right]^{1.5},\ \ \ \
\label{eq:fLCDM}
\end{eqnarray}
where where $D_{A}^{\Lambda CDM} (z)$ and $D_{A}(z)$ denote
respectively the proper angular diameter distance in the
$\Lambda$CDM reference cosmology and the current constraint model.
$A$ is the angular correction factor, which is caused by the change
in angle for the current test model $\theta_{2500}$ in comparison
with that of the reference cosmology $\theta_{2500}^{\Lambda CDM}$:
\begin{eqnarray}
&&A=\left(\frac{\theta_{2500}^{\Lambda
CDM}}{\theta_{2500}}\right)^\eta \approx
\left(\frac{H(z)D_A(z)}{[H(z)D_A(z)]^{\Lambda CDM}}\right)^\eta,
\end{eqnarray}
here, the index $\eta$ is the slope of the $f_{gas}(r/r_{2500})$
data within the radius $r_{2500}$, with the best-fit average value
$\eta=0.214\pm0.022$ \cite{ref:07060033}. And the proper (not
comoving) angular diameter distance is given by
\begin{eqnarray}
&&D_A(z)=\frac{c}{(1+z)\sqrt{|\Omega_k|}}\mathrm{sinn}[\sqrt{|\Omega_k|}\int_0^z\frac{dz'}{H(z')}].
\end{eqnarray}
It is clear that this quantity is related with $d_{L}(z)$ by
\begin{equation}
D_A(z)=\frac{d_{L}(z)}{(1+z)^2}.\nonumber
\end{equation}

In equation (\ref{eq:fLCDM}), the parameter $\gamma$ denotes
permissible departures from the assumption of hydrostatic
equilibrium, due to non-thermal pressure support; the bias factor
$b(z)= b_0(1+\alpha_b z)$ accounts for uncertainties in the cluster
depletion factor; $s(z)=s_0(1 +\alpha_s z)$ accounts for
uncertainties of the baryonic mass fraction in stars and a Gaussian
prior for $s_0$ is employed, with $s_0=(0.16\pm0.05)h_{70}^{0.5}$
\cite{ref:07060033}; the factor $K$ is used to describe the combined
effects of the residual uncertainties, such as the instrumental
calibration and certain X-ray modelling issues, and a Gaussian prior
for the 'calibration' factor is considered by $K=1.0\pm0.1$
\cite{ref:07060033}.

Following the method in Ref. \cite{ref:CBFchi21,ref:07060033} and
adopting the updated 42 observational $f_{gas}$ data in Ref.
\cite{ref:07060033}, the best fit values of the model parameters for
the X-ray gas mass fraction analysis are determined by minimizing,
\begin{eqnarray}
&&\chi^2_{CBF}=\sum_i^N\frac{[f_{gas}^{\Lambda
CDM}(z_i)-f_{gas}(z_i)]^2}{\sigma_{f_{gas}}^2(z_i)}+\frac{(s_{0}-0.16)^{2}}{0.0016^{2}}
+\frac{(K-1.0)^{2}}{0.01^{2}}+\frac{(\eta-0.214)^{2}}{0.022^{2}},\label{eq:chi2fgas}
\end{eqnarray}
where $\sigma_{f_{gas}}(z_{i})$ is the statistical uncertainties
(Table 3 of \cite{ref:07060033}). As pointed out in
\cite{ref:07060033}, the acquiescent systematic uncertainties have
been considered according to the parameters i.e. $\eta$, $b(z)$,
$s(z)$ and $K$.

\subsection{Baryon acoustic oscillation}

The baryon acoustic oscillations are detected in the clustering of
the combined 2dFGRS  and   SDSS main galaxy samples, which measure
the distance-redshift relation at $z_{BAO} = 0.2$ and $z_{BAO} =
0.35$. The observed scale of the BAO calculated from these samples,
are analyzed using estimates of the correlated errors to constrain
the form of the distance measure $D_V(z)$
\cite{ref:Okumura2007,ref:Percival2}
\begin{equation}
 D_V(z)=[(1+z)^2 D^{2}_{A}(z) \frac{cz}{H(z;p_{s})}]^{1/3}
             =H_{0}[\frac{z}{E(z;p_{s})}(\int ^{z}_{0}\frac{dz^{'}}{E(z^{'};p_{s})})^{2}]^{\frac{1}{3}}.\label{eq:DV}
\end{equation}
In this expression $E(z;p_{s})=H(z;p_{s})/H_{0}$, $D_A(z)$ is the
proper (not comoving) angular diameter distance, which has the
following relation with $d_{L}(z)$
\begin{equation}
D_A(z)=\frac{d_{L}(z)}{(1+z)^2}.\nonumber
\end{equation}
The peak positions of the BAO depend on the ratio of $D_V(z)$ to the
sound horizon size at the drag epoch (where baryons were released
from photons) $z_d$, which can be obtained by using a fitting
formula \cite{27Eisenstein}:
\begin{eqnarray}
&&z_d=\frac{1291(\Omega_{0m}h^2)^{-0.419}}{1+0.659(\Omega_{0m}h^2)^{0.828}}[1+b_1(\Omega_bh^2)^{b_2}],
\end{eqnarray}
with
\begin{eqnarray}
&&b_1=0.313(\Omega_{0m}h^2)^{-0.419}[1+0.607(\Omega_{0m}h^2)^{0.674}], \\
&&b_2=0.238(\Omega_{0m}h^2)^{0.223}.
\end{eqnarray}
In this paper, we use the data of $r_s(z_d)/D_V(z)$ extracted from
the Sloan Digitial Sky Survey (SDSS) and the Two Degree Field Galaxy
Redshift Survey (2dFGRS)
\cite{ref:Okumura2007,ref:Okumura20071,ref:Okumura20072}, which are
listed in Table \ref{baodata}, where $r_s(z)$ is the comoving sound
horizon size
\begin{eqnarray}
r_s(z)&&{=}c\int_0^t\frac{c_sdt}{a}=c\int_0^a\frac{c_sda}{a^2H}=c\int_z^\infty
dz\frac{c_s}{H(z)} \nonumber\\
&&{=}\frac{c}{\sqrt{3}}\int_{0}^{1/(1+z)}\frac{da}{a^2H(a)\sqrt{1+(3\Omega_b/(4\Omega_\gamma)a)}},
\end{eqnarray}
where $c_s$ is the sound speed of the photon$-$baryon fluid
\cite{ref:Hu1, ref:Hu2}:
\begin{eqnarray}
&&c_s^{-2}=3+\frac{4}{3}\times\frac{\rho_b(z)}{\rho_\gamma(z)}=3+\frac{4}{3}\times(\frac{\Omega_b}{\Omega_\gamma})a,
\end{eqnarray}
and here $\Omega_\gamma=2.469\times10^{-5}h^{-2}$ for
$T_{CMB}=2.725K$.

\begin{table}[htbp]
\begin{center}
\begin{tabular}{c|l}
\hline\hline
 $z$ &\ $r_s(z_d)/D_V(z)$  \\ \hline
 $0.2$ &\ $0.1905\pm0.0061$  \\ \hline
 $0.35$  &\ $0.1097\pm0.0036$  \\
\hline
\end{tabular}
\end{center}
\caption{\label{baodata} The observational $r_s(z_d)/D_V(z)$
data~\cite{ref:Percival2}.}
\end{table}
Using the data of BAO in Table \ref{baodata} and the inverse
covariance matrix $V^{-1}$ in \cite{ref:Percival2}:
\begin{eqnarray}
&&V^{-1}= \left(
\begin{array}{cc}
 30124.1 & -17226.9 \\
 -17226.9 & 86976.6
\end{array}
\right),
\end{eqnarray}
  the $\chi^2_{BAO}(p_s)$ is given as
\begin{equation}
\chi^2_{BAO}(p_s)=X^tV^{-1}X,\label{chi2-BAO}
\end{equation}
where $X$ is a column vector formed from the values of theory minus
the corresponding observational data, with
\begin{eqnarray}
&&X= \left(
\begin{array}{c}
 \frac{r_s(z_d)}{D_V(0.2)}-0.1905 \\
 \frac{r_s(z_d)}{D_V(0.35)}-0.1097
\end{array}
\right),
\end{eqnarray}
and $X^t$ denotes its transpose.

\subsection{Cosmic microwave background}

The CMB shift parameter $R$ is provided by \cite{ref:Bond1997}
\begin{equation}
 R=\sqrt{\Omega_{0m} H^2_0}(1+z_{\ast})D_A(z_{\ast})/c=\sqrt{\Omega_{m}}\int_{0}^{z_{\ast}}\frac{H_{0}dz^{'}}{H(z^{'};p_{s})},\label{R-CMB}
\end{equation}
here, the redshift $z_{\ast}$ (the decoupling epoch of photons) is
obtained using the fitting function \cite{Hu:1995uz}
\begin{equation}
z_{\ast}=1048\left[1+0.00124(\Omega_bh^2)^{-0.738}\right]\left[1+g_1(\Omega_m
h^2)^{g_2}\right],\nonumber
\end{equation}
where the functions $g_1$ and $g_2$ read
\begin{eqnarray}
g_1&=&0.0783(\Omega_bh^2)^{-0.238}\left(1+ 39.5(\Omega_bh^2)^{0.763}\right)^{-1},\nonumber \\
g_2&=&0.560\left(1+ 21.1(\Omega_bh^2)^{1.81}\right)^{-1}.\nonumber
\end{eqnarray}
In addition, the acoustic scale is related to a distance ratio,
$D_A(z)/r_s(z)$,   and at decoupling epoch it is defined as
\begin{eqnarray}
&&l_A\equiv(1+z_{\ast})\frac{\pi
D_A(z_{\ast})}{r_s(z_{\ast})},\label{la}
\end{eqnarray}
where Eq.(\ref{la}) arises a factor  $1+z_{\ast}$, because $D_A(z)$
is the proper (physical) angular diameter distance, whereas
$r_{s}(z_{\ast})$ is the comoving sound horizon. Using the data of
$l_A, R, z_\ast$ in \cite{7ywmap} and their covariance matrix of
$[l_A(z_\ast), R(z_\ast), z_\ast]$ (please see table
\ref{tab:7yearWMAPdata} and \ref{tab:7yearWMAPcovariance}), we can
calculate the likelihood $L$ as $\chi^2_{CMB}=-2\ln L$:
\begin{eqnarray}
&&\chi^2_{CMB}=\bigtriangleup d_i[Cov^{-1}(d_i,d_j)[\bigtriangleup
d_i]^t],\label{chi2-CMB}
\end{eqnarray}
where $\bigtriangleup d_i=d_i-d_i^{data}$ is a row vector, and
$d_i=(l_A, R, z_\ast)$.\\

 \begin{table}
 \begin{center}
 \begin{tabular}{c c   cc   } \hline\hline
 ~ &              7-year maximum likelihood ~~~ & error, $\sigma$ &\\ \hline
 $ l_{A}(z_{\ast})$         & 302.09      & 0.76  & \\
 $ R(z_{\ast})$             &  1.725      & 0.018 & \\
 $ z_{\ast}$                & 1091.3     & 0.91&    \\
 \hline\hline
 \end{tabular}
 \caption{The values of  $ l_{A}(z_{\ast})$, $R(z_{\ast})$, and $z_{\ast}$ from 7-year WMAP data.}\label{tab:7yearWMAPdata}
 \end{center}
 \end{table}

\begin{table}
 \begin{center}
 \begin{tabular}{c c   cc c  } \hline\hline
 ~ &             $ l_{A}(z_{\ast})$      & $ R(z_{\ast})$   & $ z_{\ast}$ &  \\ \hline
 $ l_{A}(z_{\ast})$         & 2.305      & 29.698           &  -1.333     & \\
 $ R(z_{\ast})$             &  ~         & 6825.270         &  -113.180    & \\
 $ z_{\ast}$                & ~          & ~                &  3.414      &  \\
 \hline\hline
 \end{tabular}
 \caption{The inverse covariance matrix of  $ l_{A}(z_{\ast})$, $R(z_{\ast})$, and $z_{\ast}$ from 7-year WMAP data.}\label{tab:7yearWMAPcovariance}
 \end{center}
 \end{table}

\end{document}